\documentstyle[procl]{article}

\def\Journal#1#2#3#4{{#1} {\bf #2}, #3 (#4)}

\def\NPB{{\em Nucl. Phys.} B}

\def\PLB{{\em Phys. Lett.}  B}

\def\PRD{{\em Phys. Rev.} D}

\def\be{\begin{equation}}

\def\ee{\end{equation}}

\def\bea{\begin{eqnarray}}

\def\eea{\end{eqnarray}}

\newcommand{\x}{\mbox{x}}

\begin{document}

\vspace{-4.9cm}
SHEP 96/16 \hfill \hfill \hfill \hfill July 4, 1996
\vspace{0.4cm}

\title{LATTICE DIS STRUCTURE FUNCTIONS
\footnote{Talk presented at the DIS96
International Workshop on ``Deep Inelastic Scattering and Related Phenomena'',
Rome, April 15-19, 1996}
}

\author{STEFANO CAPITANI \footnote{From October 1996 at the Desy Theory Group}}

\address{University of Southampton, Department of Physics, Highfield, \\
Southampton SO17 1BJ, England}

\maketitle\abstracts{
We present the computation, in lattice QCD, of the renormalization constants 
and mixing coefficients of operators that measure the first two moments of 
DIS Structure Functions. These calculations have been performed using the 
Sheikholeslami-Wohlert $O(a)$ improved ``clover'' action, which is known to 
reduce the systematic error associated with the finiteness of the lattice 
spacing $a$. Due to the complexities of the computations, we have developed, 
using the computer languages Schoonschip and Form, general codes that are able
to automatically carry out all the analytic lattice manipulations.}

\section{Introduction}

The computation on the lattice of renormalization constants is a necessary 
ingredient for the connection of lattice operators and matrix elements
to their continuum counterparts and the extraction of physical quantities 
  from the numbers obtained in Monte Carlo simulations.
We have computed, in lattice 1-loop perturbation theory using the  
Sheikholeslami-\-Wohlert $O(a)$ improved action, the renormalization constants 
and mixing coefficients of the quark and gluon operators of rank 
two~\cite{capitani}$^{\!-\,}$\cite{pisa} and of the quark operators 
of rank three~\cite{capitani2}$^{\!-\,}$\cite{beccarini} 
that measure the first two moments of DIS Structure Functions. 
The use of improvement~\cite{Sym}$^{\!-\,}$\cite{She} allows us to 
reduce the systematic error associated with the finiteness of the lattice
spacing $a$. In the particular formulation we have used this error 
is lowered, for on-shell quantities, from $O(a)$ to $O(a/\log a)$.

\section{Moments of Structure Functions}

The hadronic tensor $W_{\mu \nu}$, from which the Structure Functions can be 
defined in a well-known way, is written in terms of the hadronic currents as
\be
W_{\mu \nu} = \frac{1}{2 \pi} \int\! d^4 x \: e^{\displaystyle i q x} 
< p | J_{\mu}(x) J_{\nu}(0) | p >. 
\ee 
By means of a Wilson OPE expansion near the light-like region of the kind
$A(x) B(0) \sim \sum_{N,i} c_{N,i}(x^{2}) \, x^{\mu_1} \cdots 
x^{\mu_N} O^{(N,i)}_{\mu_{1} \cdots \mu_{N}}(0)$ , it can be expressed
in terms of a set of symmetric and traceless operators 
$O^{(N,i)}_{\mu_{1} \cdots \mu_{N}}$ with vanishing vacuum expectation values,
of which the dominant ones have twist two. The matrix elements of these 
operators have the general form
\be
<p| O^{(N)}_{\mu_{1} \cdots \mu_{N}} |p> = A_N(\mu) p_{\mu_{1}} \cdots
p_{\mu_{N}}  + \mbox{trace terms} \label{eq:pppp},
\ee
and are related to the moments of the Structure Functions by the 
formula
\be
\int \! d\x \: \x^{N - 1} {\cal F}_{k}(q^{2},\x)= C_{N}(q^{2} / \mu^2) 
A_{N} (\mu),
\ee
where ${\cal F}_{1} = 2 F_{1}, {\cal F}_{2} = F_{2}/\x$ and 
${\cal F}_{3} = F_{3}$. The coefficients $C_N$ are known from continuum P.T., 
and thus we can extract a given moment $<x^N>$ once we know 
the corresponding matrix element $<p| O^{(N)}_{\mu_{1} \cdots \mu_{N}} |p>$. 
Such matrix elements contain long distance (non-perturbative) physics, thus
the only viable way to compute the moments of 
Structure Functions is with the use of lattice methods. 

We have considered in our calculations the unpolarized Structure Functions, 
and in particular we have computed the operators 
\be
O^q_{\mu \nu} = \frac{1}{4} \:\overline{\psi} \:\gamma_{\{ \mu} \!
\stackrel{\displaystyle \leftrightarrow}{D}_{\nu \}} 
\psi~~~\longrightarrow~~~< \x >_q 
\ee
\be
O^g_{\mu \nu} = \sum_{\rho} \mbox{Tr} \left[ F_{\{\mu \rho} 
F_{\rho \nu\}} \right]~~~\longrightarrow~~~< \x >_g
\ee
\be 
O^q_{\mu \nu \tau} = \frac{1}{8} \,\overline{\psi} \,\gamma_{\{ \mu} \!
\stackrel{\displaystyle \leftrightarrow}{D}_{\nu} 
\stackrel{\displaystyle \leftrightarrow}{D}_{\tau \}} 
\psi~~~\longrightarrow~~~< \x^2 >_q .
\ee

\section{Improved Lattice QCD}

Lattice QCD allows the evaluation from first principles of the hadronic
matrix elements needed for the computation of the moments of the Structure 
Functions. Once discretization is introduced, the quark fields, $\psi_{n}$,
exist on the lattice sites, and the gauge fields,
$U_{n,\mu} = e^{i g_0 a t^{A} A_{n,\mu}^{A}}$, exist as links between these 
sites. The Wilson action~\cite{Wil}
\bea
S^f_W & = & a^4 \sum_{n\ } \bigg[  - \frac{1}{2 a} \sum_{\mu} \big[ 
\overline{\psi}_{n} ( r - \gamma_{\mu} ) U_{n,\mu} \psi_{n + \mu} 
+ \overline{\psi}_{n + \mu} ( r + \gamma_{\mu} ) U_{n,\mu}^{\dagger} \psi_{n} 
\big] \\
& + & \overline{\psi}_{n} \left( m_f + \frac{4 r}{a} \right) \psi_{n} \bigg]
- {\displaystyle\frac{1}{g_0^2}} \sum_{n,\mu \nu}  
\bigg[ \mbox{Tr} \left[ U_{n,\mu} U_{n + \mu , \nu} 
U_{n + \nu , \mu}^{\dagger} U_{n, \nu}^{\dagger} \right] - N_c \bigg] 
\nonumber
\eea
is widely used as discretization of the (Euclidean) QCD action. The 
corresponding regularization scheme is gauge-invariant, but the terms 
proportional to $r$ ($0 < r \le 1$), introduced to get rid of lattice 
spurious fermions, break chiral symmetry even in the case of a vanishing 
quark mass $m_f$. 

A matrix element like $<p| O^{(N)}_{\mu_{1} \cdots \mu_{N}} |p>$ (i.e. a given 
moment of the Structure Functions) can be determined from the 
computation of two- and three-point correlation functions. 
However, its determination from Monte Carlo simulations is affected by both 
statistical and systematic errors. The statistical errors come from the 
finite number of configurations used, while the systematic errors 
are of various nature: finite lattice spacing $a$, finite volume $V$, quenched
approximation (that is, dropping the contribution of the internal quark loops),
and from the necessary extrapolation to recover chiral symmetry. 

We are interested here in the systematic error associated to the finiteness 
of the lattice spacing, and to reduce this error improvement procedures 
have been proposed.\cite{Sym}$^{\!-\,}$\cite{She} The formulation that we use 
consists of adding an irrelevant operator to the standard Wilson action in 
such a way to cancel, in on-shell matrix elements,\cite{Lus} all 
terms that in the continuum limit are effectively of order ``$a$''. This 
  ``clover-leaf'' Sheikholeslami-Wohlert term is~\cite{She}
\be 
\Delta S^{f}_{I} = - i g_0 a^{4} \sum_{n,\mu \nu} \frac{r}{4 a} \: 
\overline{\psi}_{n}
\sigma_{\mu \nu} F_{n, \mu \nu} \psi_{n} \label{eq:impr} ,
\ee
where $F_{n, \mu \nu}$ is the average of the four plaquettes 
lying in the plane $\mu \nu$, stemming from the point $n$: 
\be
F_{n, \mu \nu} = \frac{1}{4} \sum_{\mu \nu = \pm} P_{n, \mu \nu} = 
\frac{1}{8ig_0a^2} \sum_{\mu \nu = \pm} (U_{n, \mu \nu} - U^{+}_{n, \mu \nu}),
\ee
with $U_{n, \mu \nu} = U_{n,\mu} U_{n + \mu , \nu} 
U_{n + \nu , \mu}^{\dagger} U_{n, \nu}^{\dagger}$. This new term in the action
 means that we have to add to the Wilson quark-quark-gluon interaction vertex,
\be
(V)^{bc}_{\rho} = -g_0 (t^A)_{bc} \left[ r \sin \frac{a(k + k')_{\rho}}
{2} + i \gamma_{\rho} \cos \frac{a(k + k')_{\rho}}{2} \right] ,
\ee
the improved quark-quark-gluon interaction vertex
\be
(V^I)^{bc}_{\rho} = -g_0 \frac{r}{2} (t^A)_{bc} 
\cos \frac{a(k - k')_{\rho}}{2} \sum_{\lambda}
\sigma_{\rho \lambda} \sin a(k - k')_{\lambda} .
\ee
The fermion propagator is not modified by this improved action, 
and neither is the gluon propagator as the first corrections to 
the pure gauge term of the Wilson action are already of order $a^2$.

As well as adding the term (\ref{eq:impr}), in the calculation of a fermionic 
Green function the fermion fields undergo the rotation 
\be  
\psi \longrightarrow \left( 1 - \frac{a r}{2} \stackrel{\displaystyle
\rightarrow }{\not\!\!{D}} \right) \psi \mbox{\ \ \ ,\ \ \ }
\overline{\psi} \longrightarrow \overline{\psi} \left( 1 + \frac{a r}{2} 
\stackrel{\displaystyle \leftarrow }{\not\!\!{D}} \right) .
\ee
This rotation combined with the use of the Sheikholeslami-Wohlert action 
has been shown to remove, from on-shell hadronic matrix elements, all terms 
that in the continuum limit are effectively of order 
$a$.\cite{She}$^{\!,\,}$\cite{rom} Using this recipe, the systematic error 
related to the lattice discretization drops from order $a$ to order $a/\log a$:
\be
\left< p \left| \widehat{\cal O}_{L} \right| p' \right>_{Monte~Carlo}=
a^{d} \left[ \left< p \left| \widehat{\cal O} \right| p' 
\right>_{phys} + O(a/\log a) \right] . 
\ee 
The magnitude of the order $a$ terms is about 20--30 \%, while
the magnitude of the order $a/\log a$ terms is about 5--10 \%. 
Therefore, with the use of Sheikholeslami-Wohlert 
improvement one can achieve a remarkable decrease of the systematic error 
coming from the finiteness of the lattice spacing.

\section{Renormalization constants} 

The renormalization constants connect the bare lattice operators, $O(a)$, 
to finite operators, $\widehat{O}(\mu)$, renormalized at a scale $\mu$:
\be
\widehat{O}^l(\mu) = Z_{lk}(\mu a) O^k(a) .
\ee
These constants are fixed in perturbation theory by the same 
renormalization conditions used in the continuum.
In the flavor Singlet case there is a mixing between quark and 
gluon operators of the same rank that have the same conserved 
quantum numbers. We can then write: 
\be
\begin{array}{c}
\widehat{O}^q = Z_{qq} O^q + Z_{qg} O^g \\
\widehat{O}^g = Z_{gq} O^q + Z_{gg} O^g ,
\end{array} 
\ee
and in this case all elements of the mixing matrix
\be
\left( \begin{array}{cc}
<q| O^q |q> & <g,\sigma| O^q |g,\sigma> \\
<q| O^g |q> & <g,\sigma| O^g |g,\sigma>  
\end{array} \right) 
\ee 
have to be computed.

To this mixing (already present in continuum QCD) the lattice formulation adds 
additional mixings, induced by the breaking of (Euclidean) Lorentz 
invariance.\cite{mix} In some cases it is possible, by a careful choice of 
the Lorentz indices, to choose operators that are multiplicatively 
renormalizable on the lattice, but the higher the moment the more complicated 
the mixing pattern is.\cite{beccarini}$^{\!,\,}$\cite{new} 
The breaking of the Lorentz invariance has also bound us to develop special 
computer routines to correctly perform the Dirac algebra on the lattice.
They are a necessary ingredient in our general codes that are able to 
automatically carry out all the stages of the algebraic manipulations from 
the elementary building blocks of each Feynman diagram.

\section{Some results}

Some simulations have been performed in the past with the unimproved 
Wilson action.\cite{new}$^{\!-\,}$\cite{old} Within errors, the results of 
these simulations are consistent with experiment. 
However, the values of the renormalization constants change non-trivially 
when they are computed in the improved theory. 
As an example we give here a selection of the results for quark operators. 
At $\beta=6 / g_0^2=6.0$ (the general dependence is 
$Z=1+\mbox{const} / \beta$), and for $r=1$, we have: 
\be
\begin{array}{rll}
\widehat{O}^q_{\{12\}} =& 1.027~O^q_{\{12\}} & \mbox{Wilson} \\
(\widehat{O}^q_{\{12\}})^{\rm{I}} 
=& 1.134~(O^q_{\{12\}})^{\rm{I}} & \mbox{Improved} \\ & & \\
\widehat{O}^q_{\{123\}} =& 1.160~O^q_{\{123\}} & \mbox{Wilson} \\
(\widehat{O}^q_{\{123\}})^{\rm{I}} =& 1.252~(O^q_{\{123\}})^{\rm{I}} 
& \mbox{Improved} \\ & & \\
\widehat{O}^q_{\rm{DIS}} =& \frac{1}{3} \left[  
1.184~O_{A} + 1.156~O_{B} \right] & \mbox{Wilson} \\
(\widehat{O}^q_{\rm{DIS}})^{\rm{I}} =& \frac{1}{3} \left[ 
1.331~(O_{A})^{I} + 1.187~(O_{B})^{\rm{I}} \right] & \mbox{Improved} ,
\end{array} 
\ee
where $O^q_{\rm{DIS}} \equiv O^q_{\{411\}} - \frac{1}{2} (O^q_{\{422\}} 
+ O^q_{\{433\}})$ (which can be writtten as one-third of the sum of the
non-symmetric operators 
$O_A \equiv O^q_{411} - \frac{1}{2} (O^q_{422} + O^q_{433})$ and \linebreak 
$O_B \equiv O^q_{141} + O^q_{114} - \frac{1}{2} (O^q_{242} + O^q_{224} 
+ O^q_{343} + O^q_{334})$) is not multiplicatively renormalizable on the 
lattice.
We see there is a great difference between the Wilson and the improved
results; in particular, the improved renormalization constants are somewhat
higher. For this reason, improved simulations using the new values of the 
renormalization constants will give better insight
into the agreement with the experimental values of the moments.

Finally, we want to mention that some results are now available also for the 
polarized Structure Functions,\cite{new} although so far limited 
to the Wilson case.

\section*{Acknowledgments}
This work has been supported by the EC Contract ERBCHBGCT940665. I would like
to thank Darren Burford for checking the manuscript.


\section*{References}
\begin{thebibliography}{99}

\bibitem{capitani} 
S.Capitani and G.C.Rossi, \Journal{\NPB}{433}{351}{1995}. 
\bibitem{capitani2}
S.Capitani, Ph.D. Thesis, Univ. of Rome ``La Sapienza'', Rome, 1994. 
\bibitem{pisa}
S.Capitani and G.C.Rossi, in {\em Proceedings of the AIHENP95 International
Workshop}, eds. B.Denby and D.Perret-Gallix 
(World Scientific, Singapore, 1995).
\bibitem{beccarini0} 
G.Beccarini, Undergraduate Thesis, Univ. of L'Aquila, L'Aquila, 1993.
\bibitem{beccarini} 
G.Beccarini, M. Bianchi, S.Capitani and G.C.Rossi, 
\Journal{\NPB}{456}{271}{1995}.
\bibitem{Sym} 
K.Symanzik, in {\em Mathematical Problems in Theoretical 
Physics}, ed. R.Schrader {\it et al.}, Springer Lecture Notes in Physics, 
vol. {\bf 153}, p. 47 (1982).
\bibitem{Lus} 
M.L\"uscher and P.Weisz, \Journal{Commun. Math. Phys.}{97}{59}{1985}.
\bibitem{She} 
B.Sheikholeslami and R.Wohlert, \Journal{\NPB}{259}{572}{1985}.
\bibitem{Wil} 
K.G.Wilson, \Journal{\PRD}{10}{2445}{1974}, and in 
{\em New Phenomena in Subnuclear Physics}, ed. A.Zichichi 
(Plenum Press, New York, 1977). 
\bibitem{rom} 
G.Heatlie, G.Martinelli, C.Pittori, G.C.Rossi and C.T.Sachrajda,
\Journal{\NPB}{352}{266}{1991}  and  
\Journal{\NPB~{\em (Proc. Suppl.)}}{17}{607}{1990};
E.Gabrielli, G.Heatlie, G.Martinelli, C.Pittori, G.C.Rossi, C.T.Sachrajda 
and A.Vladikas, \Journal{\NPB~{\em (Proc. Suppl.)}}{20}{448}{1991}; 
E.Gabrielli, G.Martinelli, C.Pittori, G.Heatlie and 
C.T.Sachrajda, \Journal{\NPB}{362}{475}{1991};
R.Frezzotti, E.Gabrielli, C.Pittori and G.C.Rossi, 
\Journal{\NPB}{373}{781}{1992}.
\bibitem{mix} 
M.Baake, B.Gem\"unden and R.Oedingen, 
\Journal{Journ. Math. Phys.}{23}{944}{1982}; 
J.Mandula, G.Zweig and J.Govaerts, \Journal{\NPB}{228}{91}{1983}.
\bibitem{new}
M.G\"ockeler, R.Horsley, E.-M.Ilgenfritz, H.Perlt, P.Rakow, G.Schierholz
and A.Schiller, \Journal{\NPB~{\em (Proc. Suppl.)}}{42}{337}{1995};
\Journal{\PRD}{53}{2317}{1996}; Preprint DESY-96-031 
({\tt hep-lat/9602029}); Preprint DESY-96-084 ({\tt hep-lat/9603006}).
\bibitem{old}
G. Martinelli and C.T.Sachrajda, \Journal{\PLB}{196}{184}{1987}; 
\Journal{\NPB}{306}{865}{1988}; \Journal{\NPB}{316}{355}{1989}.
\end{thebibliography}
\end{document}